# Reply to "On the heat transfer across a vacuum gap mediated by Casimir force"

Hao-Kun Li[1], King Yan Fong[1], Xiang Zhang[1,2*]

[1]Nanoscale Science and Engineering Center, University of California, Berkeley, CA, USA

[2]Faculties of Science and Engineering, The University of Hong Kong, Hong Kong, China

*Correspondence to: xiang@berkeley.edu

In a recent arXiv paper [1] and a related article [2], S.-A. Biehs et al. comment on our work on the heat transfer induced by Casimir force [3]. They theoretically analyze the dynamics and heat transfer properties of two coupled oscillators [1,2]. All of their derived expressions match exactly with our published analysis (Supp. Info. Sec. 1 of Ref. [3]). There is *no single expression and calculated values that is inconsistent with our earlier results*. However, S.-A. Biehs et al. interpret that the Casimir heat transfer effect was not observed in our experiment. Such a misinterpretation results from a number of flaws in their reasonings: (1) a misleading comparison between thermal radiation of bulk solids and Casimir force induced heat transfer through single phonon modes, (2) a fallacious interpretation of the heat flux measurement in our work [3], and (3) misunderstanding of the concept of mode temperature. Here, we clarify several key concepts and refute their claims.

Firstly, Ref. [1] begins with a misleading comparison between thermal radiation of bulk solids and Casimir force induced heat transfer through single phonon modes. S.-A. Biehs et al. assert that since the thermal radiative heat flux ($3.5 \times 10^{-6}$ W) only leads to a ~0.02 K change of the bulk membrane temperature, the small heat flux induced by Casimir force we measured ($6.5 \times 10^{-21}$ W) causing a large change of ~18 K in mode temperature of the membrane oscillators is an "apparent contradiction". There is nothing contradictory here; it is merely due to their misunderstanding. We stated clearly throughout the paper [3] that our experiment utilizes nanomechanical oscillators to quantify Casimir force induced heat transfer through *single phonon modes*. In contrast, the thermal radiation transfers heat to the whole bulk solid. The amount of energy to heat up a single phonon mode is of course orders of magnitude smaller than that to warm up a bulk solid. Comparing these two distinct energy scales in a single plot (Fig. 2 of Ref. [1]) is misleading. It is like comparing the energy scales of heating up a glass of water and heating up a single water molecule. In our paper (Supp Info. Sec. 2 of Ref. [3]), we have presented a detailed explanation of why thermal radiation causes a negligible effect relative to Casimir heat transfer through the analysis of the thermal time constants of single-mode and bulk processes.

Secondly, S.-A. Biehs et al.'s interpretation of the heat flux measurement is fallacious. Based on the coupled-oscillator model, they derive the expressions of the heat flux ($P_{B \to A}$) (Eq. (6) in Ref. [1]) and mode temperatures (Eq. (15) in Ref. [1]), and from there they obtain the relation between the heat flux and the mode-bath temperature difference (Eq. (16) in Ref. [1]). This approach gives the impression that the validity of Eq. (16) hinges on the coupled-oscillator model, as the authors state "without the equation for $P_{B \to A}$ the experiment could not give any number for the Casimir force driven heat flux". However, such an understanding is scientifically incorrect. In fact, the heat flux equation (Eq. (16)) is independent of the coupled-oscillator model. One can consider the dynamics of a single harmonic oscillator $\ddot{u} + 2\gamma\dot{u} + \Omega^2 u = \delta F_{th}/m + F_{ext}/m$, where $\delta F_{th}$ is the thermal force with time correlation $\langle \delta F_{th}(t) \delta F_{th}(t+\tau) \rangle = \delta(\tau) 8\gamma m k_B T$. Without the external force $F_{ext}$, the mode temperature $k_B T' = m\Omega^2 \langle u^2 \rangle$ equals to the bath temperature, i.e., $T' = T$. With an external force that acts as a feedback, $T'$ could deviate from $T$, indicating heating/cooling of the oscillator. In micro-/nano-mechanical cooling experiments, $F_{ext}$ takes the general form of $F_{ext} = Au + B\dot{u} + \delta F_{ext}$ ($A/\Omega, B \ll \Omega$ assumed), which accounts for the effects of radiation pressure [4], active feedback force [5], opto-mechanical backaction [6], sideband cooling [7], etc.

The net energy flow rate from the thermal bath to the oscillator is given by $P = 2\gamma k_B(T - T')$ (Supp. Info. Sec. 1 of Ref. [3]). This is a general relation that does not depend on the coupled-oscillator model. That is to say, our measurement of the mode-bath temperature difference is an independent measurement of heat flux.

Thirdly, S.-A. Biehs et al. also have mistaken understanding of the concept of mode temperature. On one hand, they are confused between the measurements of mode temperature and coupling constant, as they wrote "In the experiment $\langle \hat{x}_A^2 \rangle$ has been measured which is of course a mean to determine $g$" [1]. This statement is incorrect. Mode temperature (proportional to mean squared displacement) and coupling constant are two distinct physical quantities that can be measured independently. In thermomechanical noise measurements, mode temperature is obtained by the area under the noise spectra while coupling constant is obtained by the frequency splitting. The former quantifies the thermal energy of the phonon mode while the latter does not. This is the reason why a measurement of mode temperature serves as a key evidence for cooling and heating of phonon modes, as shown by abundant literature. On the other hand, they claim that "a measure of the virtual (mode) temperatures of vibrating modes which does not allow an access to the Casimir force driven heat flux" [1]. They even make a general claim that "in the strong coupling limit the widely used concept of mode temperatures loses its thermodynamic foundation and therefore cannot be employed to make a valid statement on cooling and heating" [2]. Such claims are again incorrect. As we discussed above, mode temperature directly quantifies the energy of the phonon mode and its deviation from bath temperature provides information of heat flux. It is valid regardless of the coupling strength, and it is valid when the oscillator is not in thermal equilibrium with the thermal bath. This notion is also supported by a plethora of experimental and theoretical studies of micro-/nano-mechanical cooling where strong coupling is a common scenario [6,7].

In conclusion, we have shown that S.-A. Biehs et al. s' interpretation on our paper is scientifically incorrect as it is based on their misunderstanding. Our experiment shown in Ref. [3] remains the first measurement of the Casimir heat transfer effect.